\documentclass[aps,prl,nobalancelastpage,superscriptaddress,twocolumn,longbibliography,nobibnotes]{revtex4-1}

\usepackage[utf8]{inputenc}
\usepackage[american,british]{babel}
\usepackage[T1]{fontenc}
\usepackage[pdftex]{graphicx}
\usepackage{xcolor}
\usepackage{dcolumn}
\usepackage{physics}
\usepackage{braket}
\usepackage{bm}
\usepackage{amsmath,amsthm,amssymb}
\usepackage{color}
\usepackage{verbatim}
\usepackage[normalem]{ulem}
\usepackage{dsfont}

\usepackage{hyperref}
\hypersetup{
 colorlinks=true,
 linkcolor=blue,
 anchorcolor = blue,
 citecolor = blue,
 filecolor = blue,
 urlcolor = blue
}

\newcommand{\uir}{$U_{\mathrm{i}}$ }

\newcommand{\julich}{Institute of Quantum Control (PGI-8), Forschungszentrum J\"ulich, D-52425 J\"ulich, Germany}
\newcommand{\koeln}{Institute for Theoretical Physics, University of Cologne, D-50937 K\"oln, Germany}
\newcommand{\saclay}{Universit\'e Paris-Saclay, CNRS, LPTMS, 91405, Orsay, France} 
\newcommand{\paris}{Institut Universitaire de France, 75005 Paris, France}

\begin{document}

\title{Fate of a Fractional Chern Insulator under Nonlocal Interactions in Synthetic Dimensions}

\date{\today}

\author{Patrick Liam Geraghty}
\thanks{Corresponding author: p.geraghty@fz-juelich.de}
\affiliation{\julich}
\affiliation{\koeln}

\author{Alberto Nardin}
\affiliation{\saclay}

\author{Leonardo Mazza}
\affiliation{\saclay}
\affiliation{\paris}

\author{Matteo Rizzi}
\affiliation{\julich}
\affiliation{\koeln}

\begin{abstract}

Synthetic dimensions provide a powerful route to engineer topological lattice models in ultracold atomic systems, but they contain intrinsic nonlocal interactions along the synthetic direction. 
We investigate an extended Harper–Hofstadter model subject to infinite-range column interactions that mimic this synthetic nonlocality. 
By tuning this interaction strength, we demonstrate an adiabatic evolution from a Laughlin-type bosonic fractional Chern insulator to a charge-ordered Tao–Thouless–like state without closing the many-body gap.
Along this path, the many-body Chern number and the topological entanglement entropy remain unchanged, despite a pronounced restructuring of the entanglement spectrum and the loss of robustness against local perturbations. 
This adiabatic connectivity establishes a controlled bridge between topologically ordered and effectively one-dimensional charge-ordered regimes, opening potential new avenues for state preparation.  
Our results also show that conventional topological markers may fail to diagnose the breakdown of locality-protected topological order in synthetic dimensions, and identify nonlocal interactions as a powerful knob to coherently interpolate between distinct many-body regimes.

\end{abstract}

\maketitle 

\paragraph{\textbf{Introduction ---}}
The dimensionality of a system is inherently tied to its connectivity.  
In synthetic quantum systems, these connections can be manipulated.  
Recent advances in cold-atom experiments have enabled engineered couplings between internal atomic degrees of freedom, a technique that effectively augments the system’s connectivity giving rise to additional synthetic dimensions~\cite{ogogSD,Lewenstein2012} while exploiting the tractability of low dimensional platforms. Experiments have also shown the possibility of coupling synthetic dimensions to synthetic gauge fields~\cite{Celi2014,Fallani2015,fabre2024atomic}, opening an exciting path to experimentally realize phases of matter featuring topological order (TO)~\cite{OzawaPriceReview,newSDReview}, including fractional Chern insulators (FCIs)~\cite{FCI1,FCI2,FCI3}. This enriches the opportunities in the field, beyond other notable successes in more traditional setups such as small lattices, rotating traps,and Rydberg arrays~\cite{coldatomchern,Goldman2014,Goldman2016,leonard2023realization,rotatingfermionlaughlin,semeghinispinliquid,toriccodecoldatoms}.

A prominent example of synthetic dimensions is realized using the Zeeman sublevels of magnetic atoms, such as Rubidium~\cite{Celi2014}, Ytterbium~\cite{Fallani2015}, and Erbium~\cite{weitz2020synthetic}, which can be coherently coupled via two-photon Raman transitions~\cite{TwoPhotonRamanTrans}.
The interference between the Raman beams imparts a spatially dependent phase onto the atomic wavefunction, i.e. a synthetic gauge field equivalent to the Aharonov-Bohm phase acquired by a charged particle moving in a magnetic field~\cite{dalibard2011colloquium,Linsynthgaugefields,aidelsburger2013HH,ketterleHH}.  
Utilizing these tools, experiments have identified prominent lowest-Landau level (LLL) features in these systems, employing both bosonic~\cite{ChalopinFirstQH2019,WeitzQHEdge2023,Spielman2015,Genkina2019,LiLLL,valdes2021topological} and fermionic~\cite{Fallani2015,zhou2024measuringhallvoltagehall,fallani2023hallresponse,han2019band,han2022synthetic,bouhiron2024realization} atoms. 
However, these milestone experimental achievements have thus far been limited to the study of single-particle physics.  Realizing many-body and strongly-correlated states, such as fermionic integer and bosonic fractional quantum Hall (FQH) states, requires understanding the interactions intrinsic to this synthetic set-up.

When discussing topological order, \textit{locality} is an essential notion~\cite{chen2010local,locality2,locality3,locality4}.
For instance, its celebrated robustness takes place only when we consider local perturbations~\cite{Sachdev}.
In conventional models on two-dimensional lattices, two particles occupying distant sites do not interact significantly. 
In systems with synthetic dimensions, however, atoms occupying different positions along the synthetic dimension are physically co-located on the same real-space site.
As a result, they experience anisotropic interactions~\cite{barbarino2015magnetic,barbarino2016synthetic,saito2017devil,gadway2018ints,calvanese2017laughlin,chomazinters,buser2020interacting} of strength \uir irrespective of their synthetic separation. 
Given the central role of locality in the theory of topologically ordered phases, it is not clear that this infinite-range and anisotropic interaction can allow for the existence of TO.

In this Letter, we use infinite-range anisotropic interactions as a tuning knob between a bosonic fractional Chern insulator and a strongly interacting 1D gas of bosons with an additional synthetic dimension coupled to a gauge field.  We begin by discussing a simplified model of the synthetic dimensions set-up. Within this setting, we show that the presence of nonlocal interactions drastically modify the physics of the groundstate. In particular when the physical dimension density, $\rho_{1D}=\frac{N}{L_x}$, is noninteger we show that a topologically-ordered lattice Laughlin state is adiabatically connected, via $U_{\mathrm{i}}$, to a nontopological state displaying density wave order in momentum space: a trivial Tao-Thouless state~\cite{TaoThouless}.  Several topological markers, such as the many-body Chern number~\cite{ChernNumber,HatsugaiMBCN} and the topological entanglement entropy~\cite{tee1,tee2}, are remarkably unchanged by this adiabatic passage. Yet, we prove the triviality of the state by showing that robustness to local perturbations is lost and the particle entanglement spectrum~\cite{LiHaldaneTEE,RegnaultPES2012,FCI3} is substantially reshaped. This adiabatic connection nevertheless highlights the potential relevance of the protocol for controlled preparation of topological states in finite time.

\begin{figure}[t]
    \centering
    \includegraphics[width=0.9\linewidth]{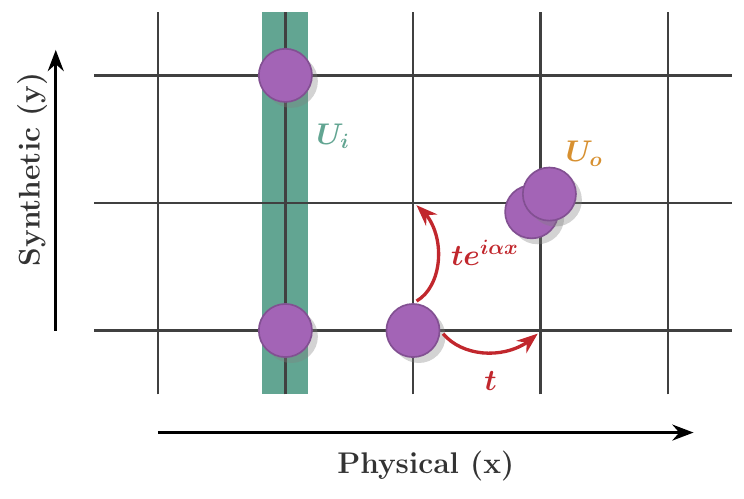}
    \caption{Schematic of the model in Eq (\ref{Eq:Ham}). The lattice has a physical (horizontal, $x$) and synthetic (vertical, $y$) direction. Hopping along the physical axis has amplitude $t$, while motion along the synthetic axis carries a Peierls phase $te^{i\alpha x}$. Interactions include an onsite term $U_{\mathrm{0}}$ and an infinite-range column interaction $U_{\mathrm{i}}$ that couples all particles within the same physical column.}
    \label{Fig:Sketch}
\end{figure}

\paragraph{\textbf{The model ---}} The model we consider is a simplified version of a synthetic dimension set-up coupled to a synthetic gauge field: essentially an extended bosonic Harper-Hofstadter model~\cite{HHModel}.  We refer to the Appendices for experimental considerations.  It contains bosons on a two-dimensional $L_x \times L_y$ square lattice pierced by a perpendicular magnetic field.  The Hamiltonian is

\begin{align}
    \hat{H} &=  - t \sum_{x,y}
    \left(
     \hat{a}_{x+1,y}^\dagger \hat{a}_{x,y}
    + e^{i \alpha x} \hat{a}_{x,y+1}^\dagger \hat{a}_{x,y} + \text{H.c.}
    \right) + \nonumber \\
    & + \frac{U_{\mathrm{0}}}{2} \sum_{x,y} \hat{n}_{x,y}(\hat{n}_{x,y}-1)
    + U_{\mathrm{i}} \sum_{x,y<y'} \hat{n}_{x,y} \hat{n}_{x, y'}.
    \label{Eq:Ham}
\end{align}
Here, $\hat{a}_{x,y}^{(\dagger)}$ are the bosonic operators which annihilate (create) a particle at the site $(x,y)$ which satisfy canonical commutation relations $[\hat{a}_{x,y}, \hat{a}_{x',y'}^\dagger] = \delta_{x,x'} \delta_{y,y'}$.  
The operator $\hat{n}_{x,y} = \hat{a}^{\dagger}_{x,y} \hat{a}_{x,y}$ is the bosonic number operator at a given site $(x,y)$. 
The term proportional to $t$ represents the kinetic energy of the model in the presence of a uniform external magnetic flux density $\alpha$.  
The model has infinite-range anisotropic interactions, proportional to $U_{\mathrm{i}}$, that describe the distinctive nonlocal nature of the synthetic dimension, here labeled by the coordinate $y$.
We parameterize the zero-range (onsite) part of this long range interaction with a second coefficient, $U_{\mathrm{0}}$ to study how the purely long-range component affects the system's phase diagram.
Throughout the Letter, we consider the hard-core limit $U_{\mathrm{0}} \to \infty$ which forbids multiple occupancy on all lattice sites.  We have also verified that taking the soft-core limit does not qualitatively alter our discussion.
To highlight some important topological features, we also consider the system to be periodic in both directions, which has recently been shown to be experimentally feasible~\cite{torusSD}.
Figure \ref{Fig:Sketch} shows a sketch of the full model.

\begin{figure}[t!]
    \centering
    \includegraphics[width=\linewidth]{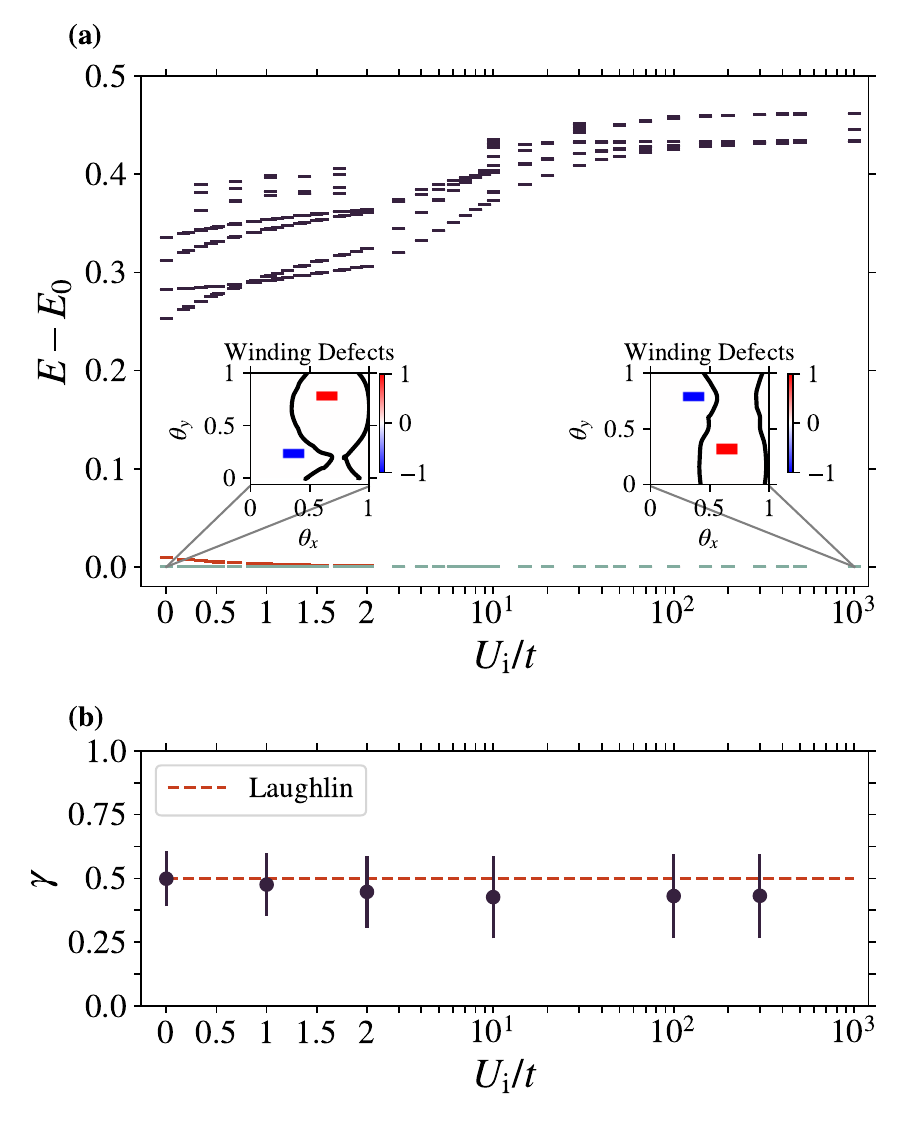}
    \caption{Analysis of some standard markers of TO between $U_{\mathrm{i}} = 0$ and $U_{\mathrm{i}}= 1000$. \textbf{(a)} the energy spectrum of an $8\times4$ lattice with $N=4$ bosons where for all values of \uir the groundstates (blue and red dashes) are degenerate and well separated from the excited states (black dashes).  The insets of \textbf{(a)} show the winding defect fields over the boundary twist angles $\theta_x$ and $\theta_y$ at $U_{\mathrm{i}}=0$ and $U_{\mathrm{i}}=1000$ where integrating over the enclosed region (black curve) returns the Chern number, $\mathcal{C}=1$ for the whole manifold; twofold degeneracy leads to $\mathcal{C}=\frac{1}{2}$~\cite{HatsugaiMBCN}.  \textbf{(b)} is the topological entanglement entropy, extracted from the linear scaling of the entanglement entropy via spatial bipartion with the bipartition perimeter as a function of \uir.  The simulation utilized TTNs for a $16\times8$ lattice with $N=8$.  It maintains a value of $\gamma \sim \frac{1}{2}$ which is the expected value for a Laughlin state at $\nu=\frac{1}{2}$ for all $U_{\mathrm{i}}$.  Detailed calculation methods are shown in the Appendices for the Chern number $\mathcal{C}$ and the topological entanglement entropy $\gamma$.}
    \label{Fig:TopoMarkers}
\end{figure}

The physics of this model is governed by three densities:
the magnetic flux density $\alpha$, the particle density $\rho_{2D}=\frac{N}{L_x L_y}$ and, noticeably, its one-dimensional counterpart, $\rho_{1D}=\frac{N}{L_x}$.
It has been previously shown in Ref.~\cite{Zeng2015ChargePumping} that at $\rho_{1D}=1$ and strong $U_{\mathrm{i}}$, a Mott-style transition takes place.
Here, we fix $\rho_{1D}=\frac{1}{2}$ and we refer to the Appendices for details of other densities $\rho_{1D} \neq \frac{1}{2}$.

\paragraph{\textbf{Topological signatures ---}}
In the presence of purely local interactions (i.e., $U_{\mathrm{i}} = 0$), previous literature has thoroughly established that a fractional Chern insulating phase emerges at a commensurate ratio of the particle and magnetic flux density of $\nu=\frac{\rho_{2D}}{\alpha}=\frac{1}{2}$~\cite{Hafezi2008,parameswaran2013fqh,MatteoFQH2017,BFQHLattice,gunnarfqh,pollmanfqh,erikFQH}.
This corresponds to a bosonic Laughlin state~\cite{ogfqhtheory} defined on a lattice, and is robust to lattice spacing effects up to $\alpha / 2 \pi \lesssim 0.4$~\cite{Hafezi2008}. We maintain this restriction in our studies as well.

We conduct our investigations by combining exact diagonalization (ED) studies on small lattices (up to $L_x = 10$) with tensor network techniques for larger lattices using a tree tensor network (TTN) approach detailed in Refs.~\cite{anthologyttn,TTNpackage,quantumtea}. The latter allows for efficient study of gapped phases of matter (among which those displaying TO~\cite{MatteoFQH2017,Macaluso2020}) in two-dimensions for sizes well beyond those accessible with ED.

We monitor three standard markers of topological order as a function of $U_\mathrm{i}$, starting from the known local limit $U_\mathrm{i}=0$.
First, the many-body spectrum in Fig.~\ref{Fig:TopoMarkers}(a)  exhibits a twofold degenerate groundstate manifold~\cite{wen_PRB_1990} (blue and red dashes), separated from excited states (black dashes) by a large energy gap, approximately $25 \times$ the groundstate splitting, Fig. \ref{Fig:TopoMarkers}(a).
Second, the state holds a nonzero many-body Chern number $\mathcal{C}=1/2$ left inset of Fig. \ref{Fig:TopoMarkers}(a).  This is extracted by integrating the winding defect field over the relevant region (black outline) and dividing by the groundstate degeneracy (d=2)~\cite{ChernNumber,HatsugaiMBCN}.
Lastly, we find a finite topological entanglement entropy~\cite{ententrop,tee1,tee2,Levin2023} (TEE) $\gamma=1/2$, Fig. \ref{Fig:TopoMarkers}(b).  The value is consistent with $\gamma=\mathrm{log}_2\mathcal{D}$, where $\mathcal D$ is the total quantum dimension: For a Laughlin state at filling $\nu=1/m$, $\mathcal{D}=\sqrt{m}$~\cite{niuHallinvariant}.  See the Appendices for details of the calculations.

Remarkably, introducing infinite-range anisotropic interactions \uir along the synthetic dimension leaves these topological signatures unchanged.  The two-fold groundstate degeneracy and large gap persist as shown in Fig. \ref{Fig:TopoMarkers}(a).  The Chern number remains $\mathcal{C}=\frac{1}{2}$ in the right inset of Fig. \ref{Fig:TopoMarkers}(a).  And in Fig. \ref{Fig:TopoMarkers}(b), the topological entanglement entropy retains the same finite value (within error bars) for all values of $U_{\mathrm{i}}$.  Taken at face value, these diagnostics would suggest that there is no phase transition and that the state at large \uir lies in the same universality class as the FCI at \uir$=0$.

\begin{figure}
    \centering
    \includegraphics[width=\linewidth]{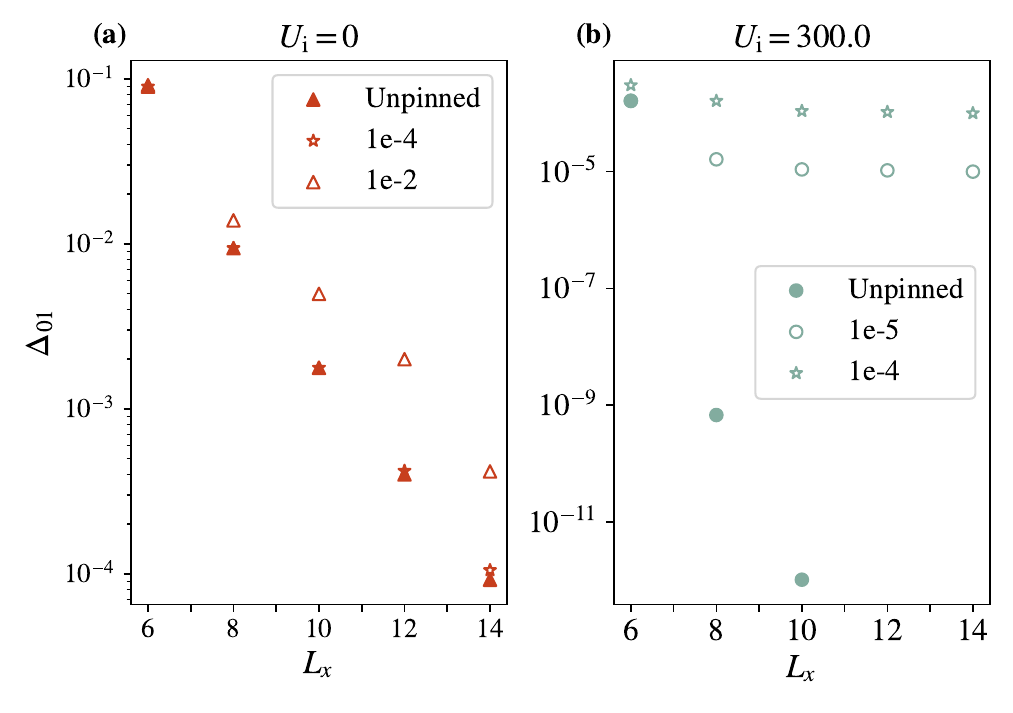}
    \caption{Topologically degenerate groundstates have a finite size splitting which decays exponentially in system size such that in the thermodynamic limit the groundstates are perfectly degenerate.  For topological states, this behavior should be robust to local pinning noise.  \textbf{(a)} shows the exponential scaling of $\Delta_{01}$ for the Laughlin case, \uir$=0$, without (filled triangles) and with (empty triangles and stars) pinning.  \textbf{(b)} is the strongly interacting case, \uir$=300$ for two other values of pinning strength $\delta$.  Without pinning, there is an exponential decay of $\Delta_{01}$, but for very weak pinning $\delta$ the splitting plateaus to a value of order $\delta$ and there is no exponential decay.  Therefore, for large \uir the groundstates are not degenerate in the thermodynamic limit and are not topologically ordered.}
    \label{Fig:Pinning:TO}
\end{figure}

\paragraph{\textbf{On the absence of topological order ---}} We now instead argue that, despite the persistence of several topological indicators, the phase emerging at large $U_{\textrm{i}}$ does not possess genuine topological order.  
The first indication in this sense comes from the robustness to local perturbations, which we test by introducing a weak single-site potential, $V(\delta)=\delta \, \hat n_{1,1}$.
Figure~\ref{Fig:Pinning:TO} shows the groundstate splitting $\Delta_{01}=E_1-E_0$ as a function of system size for a fixed aspect ratio $L_x \times L_x/2$, with $N=L_x/2$ bosons. 
In the absence of any pinning potential ($\delta=0$), for both $U_{\textrm{i}}=0$ (filled blue circles) and $U_{\textrm{i}}=300$ (filled red triangles), $\Delta_{01}$ decreases exponentially with system size, indicating that the two groundstates become degenerate in the thermodynamic limit~\cite{WenNiu1990}.  
The behavior is markedly different once the pinning is introduced (empty markers in the plot). For $U_{\textrm{i}}=0$ (red triangles and stars in the (a) panel), the splitting again closes exponentially with system size, consistent with a topologically degenerate manifold even for a relatively strong pinning potential (of order $5 \%$ of the bulk gap). 
By contrast, at large $U_{\textrm{i}}$ (blue circles and stars in the (b) panel), the splitting plateaus at a value of order $\delta$, indicating that the degeneracy is lifted in the thermodynamic limit.  The apparent degeneracy is easily broken by very weak local disorder, leading to locally distinguishable states.  We therefore conclude that the resulting state, in the thermodynamic limit where topological properties are well-defined, is not topologically ordered.

\begin{figure}[t]
    \centering
    \includegraphics[width=\linewidth]{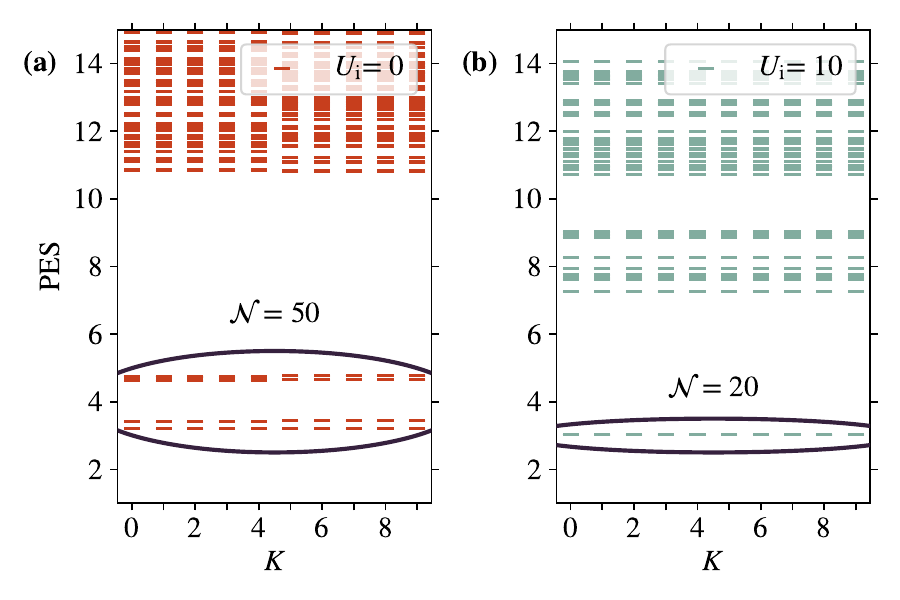}
    \caption{The momentum-resolved particle entanglement spectrum (PES) for a $10\times5$ lattice with $N=5$ using ED.  The reduced density matrix $\rho_A$ is computed in the $N_A=3$ sector.  
    Both panels display the momentum-resolved eigenvalues of $\mathcal{H}_A$ given by $\rho_A \equiv \mathrm{exp}(\mathcal{-H_A})$. \textbf{(a)} For the Laughlin state at \uir$=0$, the number of low-lying PES levels matches the expected counting $\mathcal{N}=50$.  \textbf{(b)} For strong anisotropic interactions, \uir$=10$, the PES counting changes to a value consistent with CDW structure.}
    \label{Fig:Entanglement:Spectrum}
\end{figure}

Another quantity which signals the structural modification of the ground state is the particle entanglement spectrum (PES). State counting in the PES is a long-standing method to unambiguously distinguish different topological states from trivial ones~\cite{LiHaldaneTEE,FCI3,RegnaultPES2012}, being largely insensitive to microscopic details.  
The reduced density matrix $\rho_{A}$ for  $N_A < N$ particles is constructed by tracing out a subset of the system's particles, and the entanglement Hamiltonian $H_A$ is then found by $\rho_A \equiv \textrm{exp}(-H_A)$: the PES is the full set of eigenvalues of $H_A$. 
For the bosonic Laughlin state at $\nu=\frac{1}{2}$ on a torus, the number of states in the “groundstate manifold” of the PES (i.e. the states below the entanglement gap) is $\mathcal{N} = \frac{N_{\phi}}{N_{\phi} - N_A} \binom{N_{\phi} - N_A}{N_A}$, 
where $N_{\phi}$ is the number of flux quanta piercing the torus. For a period-2 charge density wave (CDW), the PES counting reads instead $\mathcal{N}=2 \binom{N}{N_A}$.  The details of the momentum resolved PES calculation are found in the Appendices.

Figure \ref{Fig:Entanglement:Spectrum} shows the momentum-resolved PES for a $10\times5$ lattice with $N=5$ and $N_{\phi}=2N=10$ fluxes in the $N_A=3$ particles sector.  The PES ground-state manifold is in this case expected to comprise $\mathcal{N}=50$ states for Laughlin and $\mathcal{N}=20$ for CDW. 
In panel (a), the lowest part of the spectrum fits the counting, although it appears to have two branches, split into $20$ and $30$ states, respectively.
This is likely because at system size $L_x \times L_x / 2$ is slightly towards a thin-torus limit which has a preference for CDWs~\cite{TTCDW1,TTCDW2,TTCDW3,TTCDW4}.
As \uir increases, the upper branch gets pushed up into the 
continuum and only the CDW sector of $20$ states remains below the entanglement gap, see Fig.~\ref{Fig:Entanglement:Spectrum}(b) at \uir$=10$.  This is a highly nontrivial change in the entanglement properties of the ground-state manifold, and it is not accounted for in the many-body energy spectrum. It signals a drastic rearrangement of the topological structure of the state~\cite{thomale2010entanglement,Thomale2EntSpec}.

\begin{figure}[t]
    \centering
    \includegraphics[width=\linewidth]{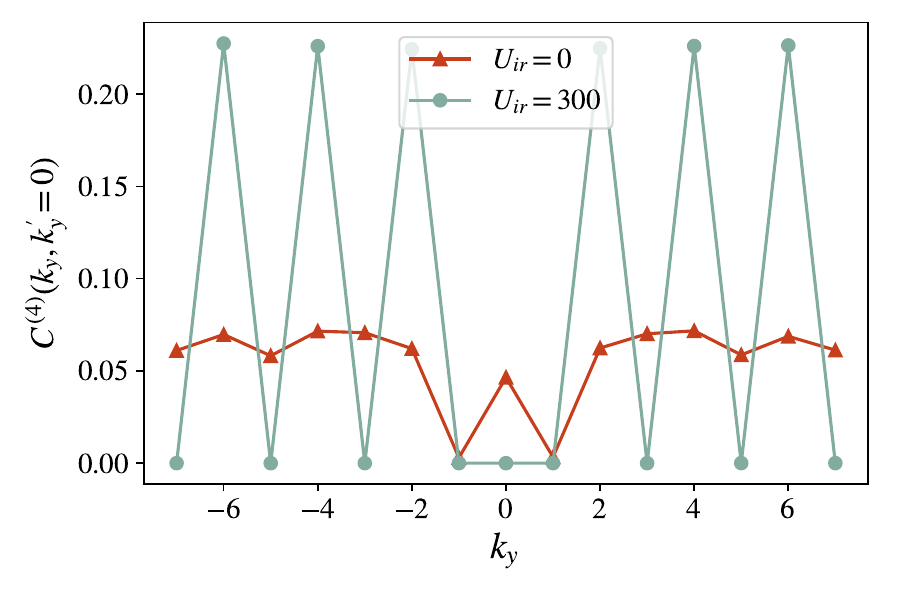}
    \caption{The four point momentum operator, Eq \ref{Eq:4PtMomentumOp}, at $k_{y}'=0$ for a $16\times8$ lattice with $N=8$, using TTNs, for both the Laughlin case with $U_{\mathrm{i}}=0$ and the strongly interacting case, $U_{\mathrm{i}}=300$.  The Laughlin case is flat and featureless in the bulk as characteristic of topological states while there is a pronounced density wave order when $U_{\mathrm{i}}=300$.}
    \label{Fig:MomentumkOrder}
\end{figure}

To further clarify the nature of this large $U_{\textrm{i}}$ phase, we first note that no real-space symmetry breaking is detected: the bulk four-point correlator remains featureless across all $U_{\textrm{i}}$ (see Appendices). 
Motivated by the structure of the Tao-Thouless thin-torus (TT) limit~\cite{TaoThouless} of fractional quantum Hall states, however, we also examine the momentum-space four-point correlator:
\begin{align}
    C^{(4)} (k_y,k_{y}') = \langle \hat{a}_{k_y}^{\dagger} \hat{a}_{k_{y}'}^{\dagger} \hat{a}_{k_{y}'}^{\phantom{\dagger}} \hat{a}_{k_{y}}^{\phantom{\dagger}}  \rangle \\
     \hat{a}_{k_{y}}^{\dagger} = \frac{1}{L_y} \sum_{x,y} \delta_{x,k_y\ell_B^2} e^{2 \pi i \frac{k_y}{L_y} y} \hat{a}_{x,y}^{\dagger}.
    \label{Eq:4PtMomentumOp}
\end{align}
The $\hat{a}_{k_y}$ are momentum-space annihilation operators associated with the lowest-Landau-level (LLL) momentum orbitals labeled by $k_y$, 
under the approximation of a very tight Gaussian
$e^{-(x-x_0)^2/2\ell_B^2}\simeq \delta_{x,x_0}$, with $x_0 = k_y \ell_B^2$, to capture the characteristic position–momentum locking of Landau-gauge LLL wavefunctions (see Appendices). As shown in Fig.~\ref{Fig:MomentumkOrder}, $C^{(4)} (k_y,k_{y}')$ is nearly uniform in the bulk for $U_{\textrm{i}}=0$, consistent with a featureless topological fluid, up to finite-size fluctuations. In stark contrast, increasing $U_{\textrm{i}}$ produces a pronounced periodic structure in momentum space, signaling the onset of charge ordering, which is consistent with the particle entanglement spectrum counting previously identified for a CDW.  We refer to this unusual trivial state as a $k-DW$.  Also, the disappearance of the peak at $(k_y,k_y')=(0,0)$ at large \uir is explained by the suppression of double occupation along a column, see Appendices for details.

We thus conclude that strong, anisotropic infinite-range interactions destabilize the Laughlin topological order and stabilize a trivial momentum-space density wave, without closing the many-body gap.

\paragraph{\textbf{Adiabatic Connection ---}}

Although the loss of topological order may appear detrimental, it opens an alternative route for state preparation. 
If two phases are connected by an adiabatic path, one can prepare the desired state through continuous parameter evolution where the time needed for preparation is $\propto 1/\Delta_{j}^{2}$, where $\Delta_j$ is the gap between the groundstate and the $j^{\mathrm{th}}$ excited state~\cite{messiah}. In conventional topological phase transitions~\cite{Sachdev}, the energy gap must close to connect to a trivial phase, preventing any truly adiabatic evolution within finite time.  In contrast, Fig~\ref{Fig:TopoMarkers}(a) shows that the many-body gap remains open and large across the transition despite connecting topological and trivial regimes. Infinite-range interactions therefore enable a topological-to-trivial transition without gap closure, implying a finite-time adiabatic path and a new route to topological state preparation.  One potential path to preparing a FQH state would begin with initializing a real-space CDW at strong $U_{\mathrm{i}}$, which hosts the same entanglement structure as Fig. \ref{Fig:Entanglement:Spectrum}(b): this can be done using an optical lattice with alternating depths~\cite{blochprepCDW,blochprepCDW2}.  In order to tune $U_{\mathrm{i}}$, a magnetic field can be used to apply a Stern-Gerlach force to spatially separate the synthetic levels~\cite{WeitzQHEdge2023,fallani2023hallresponse}.  With these two parameters as tuning knobs, there exists an adiabatic preparation path between a trivial $k-DW$ and a FCI.  Further details about the adiabatic preparation, including the dominant role of higher excited states, can be found in the Appendices.

\paragraph{\textbf{Discussion ---}} We have analyzed a minimal model of a cold atom synthetic dimension system in the bosonic Laughlin regime with infinite-range, anisotropic interactions.
At noninteger one-dimensional densities, $\rho_{1D}$, and for strong nonlocal coupling $U_{\mathrm{i}}$, topological markers remain nontrivial and bulk local operators in real space stay featureless, hinting at topological order.  
Meanwhile, robustness to local perturbations is lost and the particle entanglement spectrum shows trivial counting, indicating a topologically trivial phase. 
Taken together, these signatures reveal a hidden transition to a trivial momentum-space density-wave ($k-DW$) state, occurring without closure of the many-body gap.
The degree of nonlocality required for the interactions to drive this type of transition is a subject of future work.

The peculiar behaviors we highlight here are a challenge to the definition of TO as it is commonly understood.  
They originate from the infinite-range interaction proportional to $U_{\mathrm{i}}$, which implements a genuinely nonlocal manipulation of the state: an ingredient at odds with the locality assumptions that underlie much of TO theory.

At the same time, we have demonstrated the possibility of topologically trivial states that nevertheless exhibit nonzero topological entanglement entropy and a nonzero Chern number. 
In this sense, synthetic dimensions provide an exceptionally controlled platform for stress-testing the foundations of TO and for disentangling which seemingly topological diagnostics remain robust in the presence of engineered nonlocality.

\paragraph{\textbf{Note ---}} 
During the preparation of this Letter, we became aware of the related preprint by D. G. Reid \textit{et al}. (arXiv:2602.21108), which investigates interacting phases of a Harper–Hofstadter model motivated by synthetic dimensions constructed from harmonic trap states. While both works address interaction effects in synthetic-dimension realizations of Harper–Hofstadter physics, the nature of the effective interactions and the parameter regimes explored are substantially different from those considered here.

\paragraph{\textbf{Acknowledgements ---}} We are grateful to Niklas Tausendpfund, Erik Weerda, Nathan Goldman, Cécile Repellin, Raphael Lopes, Felix Palm, Sylvain Nascimbene and Martin Weitz for discussions on synthetic dimensions and collaborations on related subjects.  The Tree Tensor Network package TTN.jl~\cite{TTNpackage} was used for large scale tensor network simulations and the authors warmly thank Niklas Tausendpfund and Noah Elbracht for its development and improvement. P.L.G. and M.R. acknowledge the support from the DFG under Germany’s Excellence Strategy - Cluster of Excellence Matter and
Light for Quantum Computing (ML4Q) EXC 2004/1-390534769 and the project by the DFG Collaborative Research Center (CRC) 183 Project No. 27710199.  The authors gratefully acknowledge the Gauss Centre for Supercomputing e.V. (www.gauss-centre.eu) for funding this project by providing computing time through the John von Neumann Institute for Computing (NIC) on the GCS Supercomputer JUWELS~\cite{juwels} (Grant NeTeNeSyQuMa) and the FZ Jülich for computing time on JURECA~\cite{jureca} (institute project PGI-8) at Jülich Supercomputing Centre (JSC).  A.N. and L.M. acknowledge this work is part of HQI (www.hqi.fr) initiative and is supported by France 2030 under the French National Research Agency grant number ANR-22-PNCQ-0002.

\bibliographystyle{apsrev4-2}
\bibliography{references}

\clearpage
\onecolumngrid

\begin{center}
    \Large\bfseries Appendices
\end{center}

\vspace{1em}

\twocolumngrid

\begin{figure*}[t!]
    \centering
    \includegraphics[width=\textwidth]{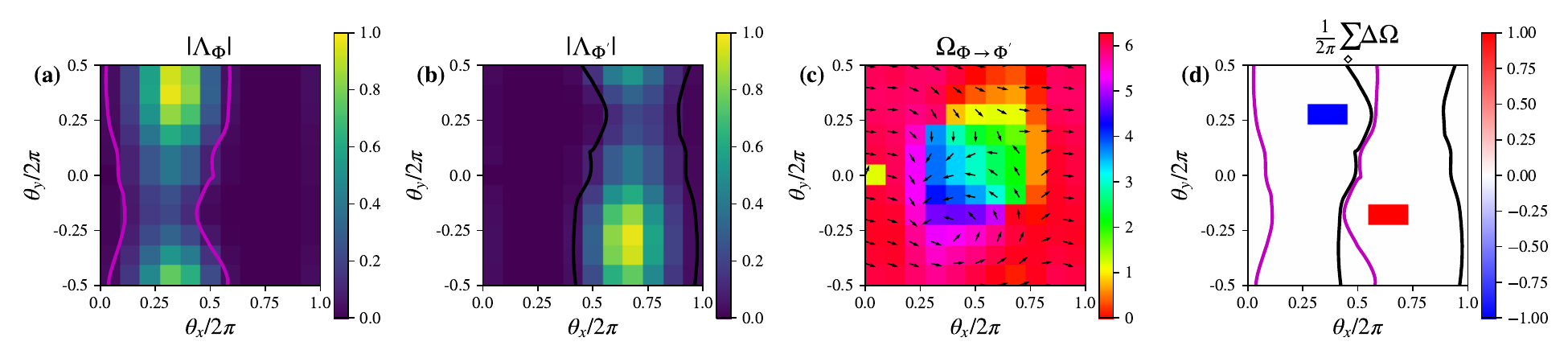}
    \caption{The sequence of figures here show the steps of calculating the many-body Chern number using the Hatsugai method~\cite{HatsugaiMBCN} for an $8\times4$ lattice with $N=4$.  \textbf{(a)} and \textbf{(b)} show the determinant of $\Lambda_{\Phi}$ and $\Lambda_{\Phi^{'}}$ and the two complementary regions of each reference multiplet, the black and purple outlines.  \textbf{(c)} shows the phase $\Omega_{\Phi \rightarrow \Phi^{'}}$ which contains nontrivial vortices. And finally \textbf{(d)} shows the winding defects where integrating over the relevant region returns the many-body Chern number.}
    \label{Fig:EndMatter:HatsugaiSteps}
\end{figure*}

\paragraph{\textbf{Chern number calculation ---}} We compute the many-body Chern number using the overlap-based method of Hatsugai~\cite{HatsugaiMBCN}, which evaluates the topology of a degenerate groundstate manifold under twisted toroidal boundary conditions.  Figure \ref{Fig:EndMatter:HatsugaiSteps} shows the sequence of steps to calculate the Chern number.  For each twist angle in $(\theta_x,\theta_y) \in [0,1] \times [0,1]$, we obtain the two groundstates forming the manifold. The twists are implemented by modifying the boundary-crossing hoppings:
\begin{align}
    \hat{a}_{1,y}^\dagger \hat{a}_{L_x,y} \rightarrow e^{2\pi i \theta_x} \hat{a}_{1,y}^\dagger \hat{a}_{L_x,y} \\
     \hat{a}_{x,1}^\dagger \hat{a}_{x,L_y} \rightarrow e^{2\pi i \theta_y} \hat{a}_{x,1}^\dagger \hat{a}_{x,L_y}.
    \label{Eq:EndMatter:TwistedBoundaryHopping}
\end{align}

Because the phase is insulating, the many-body gap remains open for all twist angles meaning only the groundstate manifold is energetically relevant.  To evaluate the Chern number, we choose two reference multiplets, $\Phi(\Theta_x,\Theta_y)$ and $\Phi'(\Theta_x',\Theta_y')$, each containing the two degenerate groundstates but obtained at twist angles far from each other to minimize correlations. The groundstate projector is defined as
 \begin{align}
    P(\theta_x,\theta_y) = |\Psi_0(\theta_x,\theta_y)\rangle\langle\Psi_0(\theta_x,\theta_y)| + \\
     |\Psi_1(\theta_x,\theta_y)\rangle\langle\Psi_1(\theta_x,\theta_y)|
     \label{Eq:Endmatter:GSProjector}
 \end{align}
 
Using this projector, we build a $4\times4$ matrix composed of four $2\times2$ blocks defined by:
\begin{align}
    \Lambda_{\Phi^{(')}}(\theta_x,\theta_y) = \langle \Phi^{(')}_j | P(\theta_x,\theta_y) | \Phi^{(')}_k \rangle
\end{align}
with $j,k = 0,1$.  Each of the four blocks corresponds to each pairing of the two reference multiplets $\Phi$ and $\Phi'$.  The determinants of the two diagonal blocks are shown in Fig.~\ref{Fig:EndMatter:HatsugaiSteps}(a–b).  The regions of relevance for each multiplet form two complementary regions: the black and purple outlines.  The phase field, related to the Berry curvature, is obtained from the phase of the determinant of one of the off-diagonal blocks:
\begin{align}
\Omega_{\Phi \rightarrow \Phi'}(\theta_x,\theta_y) = \arg[\det(\langle \Phi_j | P(\theta_x,\theta_y) | \Phi'_k \rangle )],
\end{align}
as shown in Fig. \ref{Fig:EndMatter:HatsugaiSteps}(c).  The many-body Chern number, $\mathcal{C}$, is determined by counting the number of vortices (with sign) contained within the region complementary to the reference-multiplet patches, and dividing by the groundstate degeneracy.  Figure~\ref{Fig:EndMatter:HatsugaiSteps}(d) is the vortex defect field; a single vortex appears in this region, which, combined with the two-fold degeneracy, gives $\mathcal{C}$ per sector of $\frac{1}{2}$.

\paragraph{\textbf{Topological Entanglement Entropy ---}}
For gapped ground states, the bipartite entanglement entropy obeys a boundary law, $S(A)=\alpha|\partial A|-\gamma+\cdots$, where $\gamma$ is the universal topological entanglement entropy~\cite{ententrop,tee1,tee2}. Figure~\ref{Fig:TopoMarkers}(b) shows that $\gamma$ extracted from our TTN data remains approximately constant as \uir is varied, which naively suggests a stable topological phase. We obtain $\gamma$ by bipartitioning the TTN at successive hierarchical layers; for the $16\times8$ system the TTN has seven layers, and we use cuts from layers $3, 4, 5, 6\textrm{, and }7$ with perimeters $6,8,12,16,$ and $24$ respectively, with a linear fit and identifying $\gamma$ with the negative $y$-intercept.

Recent results~\cite{Levin2023,Levin2024} show that $\gamma$ actually provides a lower bound on subleading corrections to the boundary law.
Consistent with this constraint, the apparent decrease of $\gamma$ at large \uir in Fig.~\ref{Fig:TopoMarkers}(b) could be interpreted as contamination from nonuniversal finite-size corrections rather than a change in the universal TEE: a $\gamma$ that is positive but less than the expected universal value cannot by itself establish $\gamma\neq0$. 
Within our uncertainties, the extracted value remains compatible with $\gamma\simeq 1/2$, and the downward drift at large \uir may simply come from the fit uncertainties seen in the large error bars.

\paragraph{\textbf{Particle Entanglement Spectrum ---}} 
Here we briefly review the calculation of the Particle Entanglement Spectrum (PES), first introduced in the FQHE context in Refs.~\cite{Haque_PRL_2007, Zozulya_PRB_2007,Sterdyniak_PRL_2011}.

PES works by partitioning the quantum state into two sets of particles; this is formally achieved by introducing a fictitious spin-$1/2$ degree of freedom~\cite{Dubail_PRB_2012} for each particle, $\hat a_{x,y}\rightarrow (\hat a_{x,y}^A+\hat a_{x,y}^B)/\sqrt{2}$,
projecting onto fixed particle-number sectors $(N_A,N_B=N-N_A)$ and tracing out the pseudospin-$B$ degrees of freedom: $\hat\rho_A = {\rm Tr}_B\hat \rho$.
Such a bipartition has the advantage of preserving the geometry (the number of orbitals) of the system~\cite{regnault2015_review}, and, in the case of a system with periodic boundaries, also translation invariance: 
if the ground-state has a total momentum $\bf K$, then ${\bf K}_A+{\bf K}_B=\bf K$.

More formally, we can decompose the ground-state wavefunction as
\begin{equation}
    \ket{\text{GS}({\bf K})} = \sum_{\mu_A, \mu_B} \mathcal P^{}_{\mu_A,\mu_B} \ket{\mu_A} \otimes \ket{\mu_B},
\end{equation}
where 
\begin{equation}
    \begin{split}
        \ket{\mu_A} &= \ket{N_A,{\bf K}_A, a},\\
        \ket{\mu_B} &= \ket{N_B=N-N_A,{\bf K}_B={\bf K}-{\bf K}_A,b}
    \end{split}
\end{equation}
and $a,b$ label additional quantum numbers within a fixed particle number and momentum state. After tracing out the pseudospin-$B$ degrees of freedom the reduced density matrix is block-diagonal in each $(N_A, \mathbf{K}_A)$ sector due to momentum and particle-number conservation; within each block one has
    $\rho_A = \mathcal P\mathcal P^\dagger$
so that the entanglement Hamiltonian reads
    $\mathcal H_{\rm PES} = -\log \rho_A$.
If ground-state degeneracy occurs, it is known that the density matrix for the PES should account for all the $d$ degenerate states $\ket{\Psi_r}$ by incoherently summing their density matrices~\cite{Sterdyniak_PRL_2011,FCI3,regnault2015_review}: $\hat\rho = \frac{1}{d}\sum_{r=1}^d\ket{\Psi_r}\bra{\Psi_r}$.
With the same decomposition above, the entanglement Hamiltonian reads
\begin{equation}
    \mathcal H_{\rm PES} = -\log\left(\frac 1d \sum_{r=1}^d\mathcal P_r^{}\mathcal P^\dagger_r\right).
\end{equation}

Finally we briefly review the expected universal counting of the entanglement energies below the entanglement gap for a fractional Chern insulator and a charge density wave~\cite{bernevig2012thintorus,RegnaultPES2012}.
In the CDW case, each of the two ground states is, in momentum space, a product state.
The counting of the entanglement energies for each such state is therefore given by the number of ways in which we can select $N_B = N-N_A$ particles out of $N$, i.e. $\binom{N}{N-N_A}=\binom{N}{N_A}$. In total we therefore have
\begin{equation}
    \mathcal N_{\rm CDW} = 2\binom{N}{N_A}
    \label{Eq:PEScountingCDW}
\end{equation}
states below the entanglement gap.
On the other hand in the FCI case the counting is significantly larger, due to the correlated nature of the ground states, but still much smaller than the total Hilbert space dimension. 
It can be obtained by a fractional-counting statistics~\cite{haldane_PRL_1991} argument, since in a Laughlin phase the zero modes satisfy a generalized Pauli exclusion principle~\cite{bernevig_PRL_2008}: in our case, this disallows having more than $1$ particle in $2$ consecutive orbitals.
The counting can therefore be obtained by a purely combinatorial analysis, counting binary strings of length $N_\phi$ with $N_A$ ones and no adjacent ones (with periodic boundary conditions). The result is
\begin{equation}
    \mathcal N_{\rm FCI} = \frac{2N}{N_A!}\frac{(2N-(N_A+1))!}{(2N-2N_A)!}= \frac{N_\phi}{N_\phi-N_A}\binom{N_\phi-N_A}{N_A},
    \label{Eq:PEScountingFCI}
\end{equation}
which is the one quoted in the main text.

\begin{figure}
    \centering
    \includegraphics[width=0.9\linewidth]{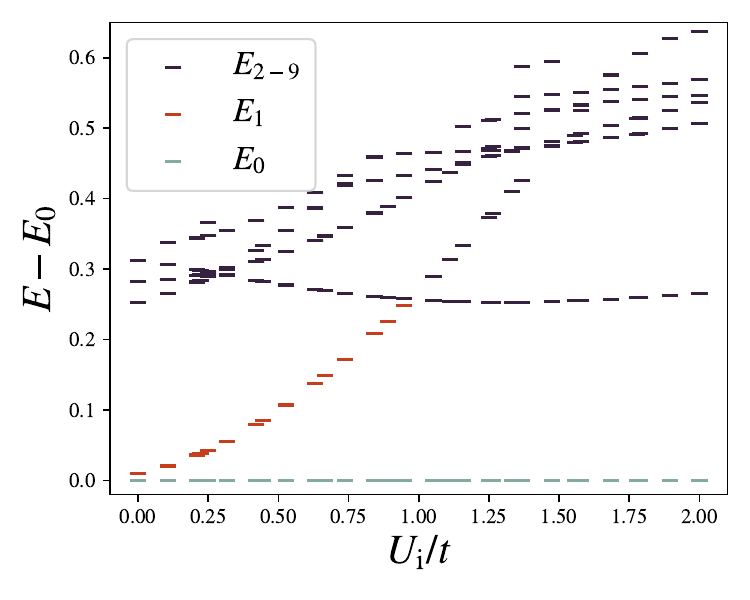}
    \caption{The energy spectrum as a function of $U_{\mathrm{i}}$ for a $4\times8$ lattice with $N=4$ and thus $\rho_{1D}=1.0$.  As the nonlocal interaction strength grows, the topological two-fold degeneracy at $U_{\mathrm{i}}=0$ breaks and one of the groundstates mixes with the spectrum resulting in a trivial CDW phase.}
    \label{Fig:EndMatter:Spectrum-rho1d-1p0}
\end{figure}

\paragraph{\textbf{Dependence on $\rho_{1D}$ ---}} As shown previously by Zeng,  Wang, and Zhai in 2015~\cite{Zeng2015ChargePumping}, the parameter that controls the topological character of a model with anisotropic synthetic dimension interactions is the physical dimension density, $\rho_{1D}=\frac{N}{L_x}$.

\begin{figure}
    \centering
    \includegraphics[width=\linewidth]{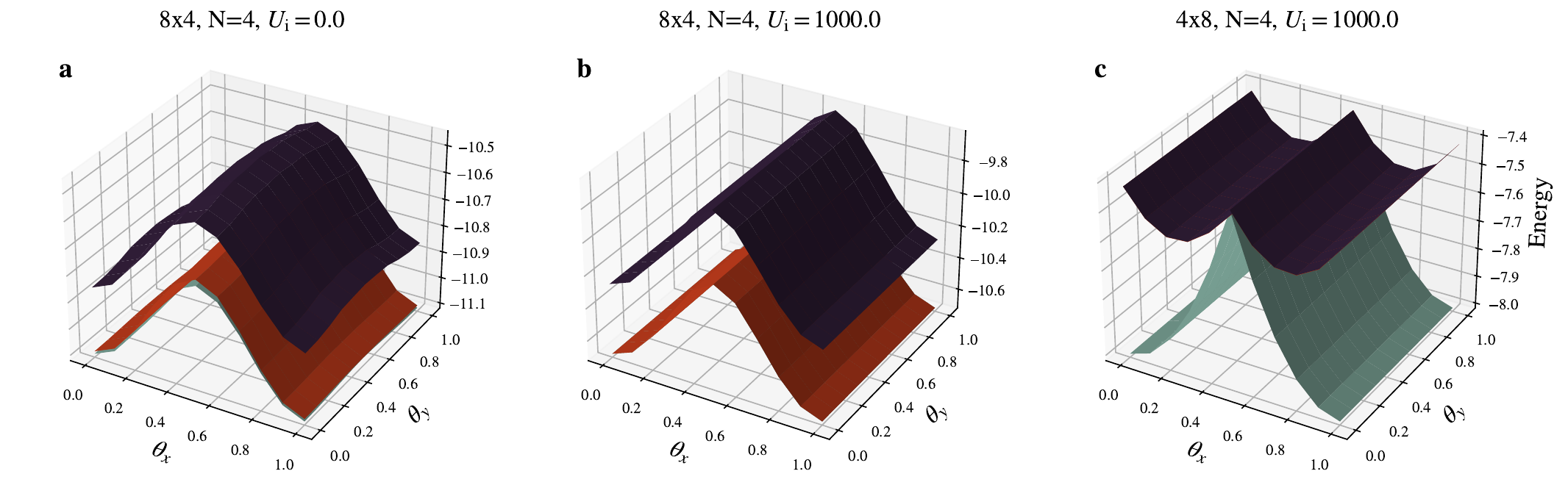}
    \caption{When applying twisted boundary conditions, in order to calculate a nontrivial Chern number, the state must be insulating: no mixing between the groundstate manifold and higher states.  Plots \textbf{(a)} and \textbf{(b)} show the spectrum under twisting for an $8\times4$ lattice with $N=4$ at \uir$=0$ and \uir$=1000$, respectively.  Both show insulating behavior which host nontrivial Chern number, however, this is not the case for \textbf{(c)} which is a $4\times8$ lattice with $N=4$.  These plots display the importance of $\rho_{1D}=\frac{N}{L_x}$.}
    \label{Fig:SuppMat:SpectrumUnderTwisting}
\end{figure}

To illustrate this dependence, in Fig.~\ref{Fig:EndMatter:Spectrum-rho1d-1p0} we consider a $4 \times 8$ lattice with $N=4$, and thus $\rho_{1D}=1$, which has the same flux density $\alpha=1/4$ as the $8\times4$ case shown in Fig. \ref{Fig:TopoMarkers}.  
Although the two-fold Laughlin degeneracy at $U_{\mathrm{i}}=0$ is protected by a sizable gap, even a small nonzero $U_{\mathrm{i}}$ splits the groundstate manifold and one of the original ground states is rapidly absorbed into the spectrum. Under boundary-condition twists, many higher states then mix with the ground state, eliminating the spectral gap and displaying a topologically trivial, noninsulating regime, see Fig. \ref{Fig:SuppMat:SpectrumUnderTwisting}(c).

In contrast, it was demonstrated above that at incommensurate $\rho_{1D}=\frac{1}{2}$, the two-fold degeneracy protected by a large gap is maintained for all values of $U_\mathrm{i}$, see Fig. \ref{Fig:TopoMarkers}(a).
Additional spectra confirm that this robustness also holds for $\rho_{1D}=\frac{1}{3}$ and $\frac{2}{3}$.  
Together these results align with Ref.~\cite{Zeng2015ChargePumping}: at commensurate $\rho_{1D}$, the topological response becomes trivial, whereas for all other incommensurate $\rho_{1D}$, the system retains a nontrivial topological marker.

\paragraph{\textbf{Four point momentum projection ---}} The anisotropic nature of the synthetic dimension interactions leads us to think about the strongly interacting state as a type of Tao-Thouless state~\cite{TaoThouless} which is a wavefunction ansatz for the Laughlin state in the thin-torus limit.  The state is characterized by single-particle orbitals in the lowest Landau level (LLL) given by:
\begin{align}
\phi_m (x,y) \propto e^{- \frac{(x-k_m l_{B}^2)}{2 l_{B}^2}} e^{ik_m y}
\label{Eq:EndMatter:TaoThoulessOrbitals}
\end{align}
where $l_B$ is the magnetic length and $k_m = 2 \pi m / L_y$ are the discrete $y$-direction momenta.  These orbitals are wavepackets localized along the $x$-direction and plane waves along $y$: the same effect as the nonlocal interaction.  By transforming the real-space four-point operator in the manner described by Eq.~\eqref{Eq:4PtMomentumOp}, we are projecting down to the LLL in the Tao-Thouless regime.

\begin{figure}[h]
    \centering
    \includegraphics[width=\linewidth]{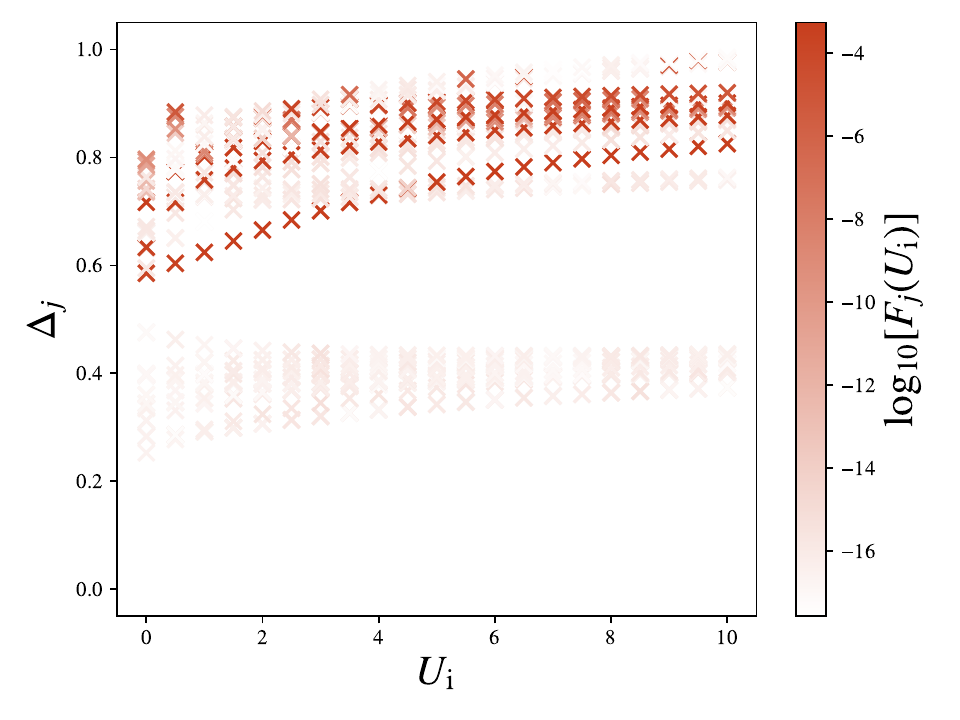}
    \caption{The energy spectrum as a function of the infinite-range interaction strength $U_{\mathrm{i}}$ where the color corresponds to the log of the adiabatic condition term between the groundstate manifold and the $j^{th}$ excited state.  The nearest excited states, in fact, play a sub-dominant role to the adiabatic transfer time; higher excited states are orders of magnitude more relevant.}
    \label{Fig:AdiabaticCondition:SpectrumUir}
\end{figure}

\paragraph{\textbf{High excited states and adiabatic time ---}} If two phases are connected by an adiabatic path, one can prepare the desired state through continuous parameter evolution governed by the adiabaticity parameter: $F_{j}(\beta) = \big|\langle \Psi_0 \big| \frac{\partial \hat{H}}{\partial \beta} \big| \Psi_{j} \rangle\big| / \Delta_{j}^{2}$, where $\beta$ is a tunable parameter and $\Delta_j$ the excitation gap to the $j^{\mathrm{th}}$ state. 
The total time for adiabatic preparation along a path $C$ is $T^{(\beta)} = \max_j \big ( \int_C F_{j}(\beta) d\beta \big )$. 
However, when the gap closes, $F_{j}(\beta)$ diverges, and the time needed to traverse that gap closure point adiabatically becomes infinite~\cite{messiah}.
In order to study any adiabatic path using the adiabaticity parameter, it is crucial to not only consider the gap magnitude; it does not alone set the minimal time scale for adiabatic evolution.  Physically, the matrix element encodes the susceptibility of the ground state to transitions into $|\Psi_j\rangle$ under variations of the control parameter $\beta$.  Figure \ref{Fig:AdiabaticCondition:SpectrumUir} shows the log-weight of the adiabaticity parameter for excited states up to $N=50$.  The low-lying excited states are practically negligible to the total preparation time.  Instead, it is the higher states which contribute multiple orders of magnitude stronger.  Therefore, the energy gap and the low-lying states alone are not enough; higher energy level matrix elements are essential to characterize the adiabatic path.

\paragraph{\textbf{Approximations in the Hamiltonian ---}}  Now we consider the limitations of the simplified model, Eq.~\eqref{Eq:Ham}, approximating a synthetic dimensions experiment.  The $y$-direction hopping amplitude in the simplified model is realized in a synthetic dimensions set-up via a two-photon Raman transition~\cite{SynthDimHopping1,SynthDimHopping2}.  A fully accurate simulation would also need to take into account the inhomogeneities in the hopping rates which will depend on the synthetic levels involved in the transition~\cite{Celi2014}.

Not only the hopping amplitudes depend on the synthetic level, but also the interactions between the different levels.  In our simplified model, we take the interactions between all levels to all be equal at \uir; this is the same simplification as Ref.~\cite{Zeng2015ChargePumping}.  However, the inter-level interactions are not necessarily all the same as discussed in the experimental Ref.~\cite{DalibardScatteringAmps}.

\begin{figure}
    \centering
    \includegraphics[width=0.5\linewidth]{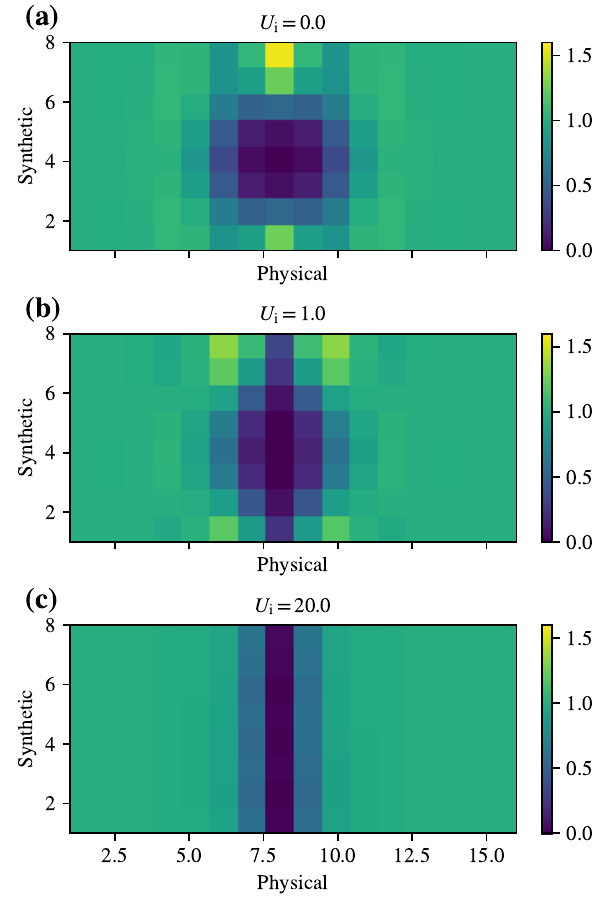}
    \caption{These plots show the real space normalized density-density correlation, Eq \ref{Eq:SuppMat:RealSpaceFourPoint}, for a $16\times8$ lattice with $(x_0,y_0) = (8,4)$.  In the Laughlin case \textbf{(a)}, there is the expected correlation hole where the radius is the commonly known screening radius.  For larger values of \uir, the anisotropic nature of the nonlocal interaction causes the correlation hole to extend along the synthetic dimension, correlating the entire column.}
    \label{Fig:SuppMat:fourpt-realspace}
\end{figure}

\paragraph{\textbf{Real space four-point correlator ---}} For two topologically degenerate states, no local operator exhibits symmetry-breaking expectation values which can distinguish the two degenerate states: local probes are featureless~\cite{chen2010local,locality2,locality3,locality4}.  The operator we study is the normalized density-density correlator in real space

\begin{equation}
    F(x,y) = \frac{\langle a_{x_0,y_0}^{\dagger} a_{x,y}^{\dagger} a_{x,y} a_{x_0,y_0} \rangle}{\langle n_{x_0,y_0} \rangle \langle n_{x,y} \rangle}
    \label{Eq:SuppMat:RealSpaceFourPoint}
\end{equation}

where Figure \ref{Fig:SuppMat:fourpt-realspace} takes $(x_0,y_0) = (8,4)$ for a $16\times8$ lattice with $N=8$.  For the Laughlin-type state at $U_{\mathrm{i}}=0$, Fig \ref{Fig:SuppMat:fourpt-realspace}(a), there is a correlation hole centered at $(x_0,y_0)$ where the radius is the screening radius.  All sites outside the screening radius are uncorrelated with $(x_0,y_0)$; this is a well-known behavior of quantum Hall liquids~\cite{screeninglength}.

As the nonlocal interaction strength, $U_{\mathrm{i}}$, increases, the correlation hole becomes extended along the synthetic dimension as the anisotropic interactions correlate the whole column.  However, the bulk region of the lattice still remains entire uncorrelated: local probes are featureless.  This is another characteristic property of a topological state that the strong \uir state holds even though we know it is not topologically ordered.

\paragraph{\textbf{Behavior of $C^{(4)}(k_y,k_y')$ at $(0,0)$ ---}}
For the Laughlin case at $U_{\mathrm{i}}=0$, the four-point momentum-space correlator exhibits a pronounced peak at $(k,k')=(0,0)$. In this limit Eq.~\ref{Eq:4PtMomentumOp} reduces to

\begin{equation}
C^{(4)}_{0,0} = \langle \hat{a}_{0}^{\dagger} \hat{a}_{0}^{\dagger} \hat{a}_{0} \hat{a}_{0} \rangle
= \frac{1}{L_{y}^{4}} \sum_{y,y',y'',y'''} 
\langle \hat{a}_{0,y}^{\dagger} \hat{a}_{0,y'}^{\dagger} 
\hat{a}_{0,y''} \hat{a}_{0,y'''} \rangle .
\end{equation}

This expression sums over all pair-hopping processes along the $y$ direction within the first column. In the interacting case ($U_{\mathrm{i}}\neq0$), double occupation of a physical site (i.e. a column) is suppressed, and such processes are therefore strongly reduced. In contrast, when $U_{\mathrm{i}}=0$ two particles can occupy the same column, allowing pair-hopping events along the vertical direction and generating a finite contribution to $C^{(4)}(0,0)$. The resulting enhancement produces the peak observed at $(k,k')=(0,0)$. Notably, this mechanism is distinct from the real-space correlation hole discussed above.

\paragraph{\textbf{Spectral behavior under twisting ---}}  In order to calculate the many-body Chern number using Hatsugai's method~\cite{HatsugaiMBCN}, it is required that the groundstate manifold does not mix with other states under boundary condition twisting. Figure \ref{Fig:SuppMat:SpectrumUnderTwisting}(a) shows the three lowest-lying states as a function of the twist angles $\theta_x$ and $\theta_y$.  The groundstate is two-fold degenerate and the first excited state stays well separated for all values of twist angle.  This spectral behavior is also seen in Figure \ref{Fig:SuppMat:SpectrumUnderTwisting}(b) which is at \uir$=1000$; both plots show a regime with nontrivial Chern number.

Figure \ref{Fig:SuppMat:SpectrumUnderTwisting}(c) shows the spectrum under twisting for a $4\times8$ lattice with $N=4$ which means $\rho_{1D}=\frac{N}{L_x}=1.0$.  In this case, higher states mix with the groundstate manifold meaning that the state is not insulating and therefore cannot host a nontrivial Chern number: a trivial CDW state.  The difference between Figures \ref{Fig:SuppMat:SpectrumUnderTwisting}(b) and (c) display the importance of $\rho_{1D}$.  With the same flux density and number of sites, commensurate values of $\rho_{1D}$ lead to noninsulating behavior at strong \uir, while incommensurate $\rho_{1D}$ maintains the insulating spectrum.

\paragraph{\textbf{Phase transitions beyond conventional criticality ---}} Our results sit at the intersection of two complementary insights about fractional phases. On the one hand, the nonlocal interaction along the synthetic dimension imposes strong kinematic constraints that stabilize fractional groundstate degeneracy and a quantized many-body Chern number.  This is reminiscent of center-of-mass conserving constructions, where degeneracy is enforced algebraically rather than by symmetry breaking~\cite{moore2005}. On the other hand, the adiabatic path we identify explicitly shows that such seemingly topological invariants can persist even as the state evolves into a locally ordered configuration that is topologically trivial, induced by violating the locality assumptions underlying the topological protection~\cite{transwoclosing2013}. In our case, the nonlocal synthetic interaction changes the structural assumptions defining the classification of the phase, allowing an adiabatic interpolation without bulk gap closing while also reshaping the entanglement spectrum and eliminating robustness to local perturbations. The resulting picture is that fractional degeneracy and quantized response can survive beyond the regime of genuine topological order, and that synthetic dimension interactions allow for a concrete microscopic setting in which topological order can be unwound without conventional criticality.

\begin{figure}
    \centering
    \includegraphics[width=0.8\linewidth]{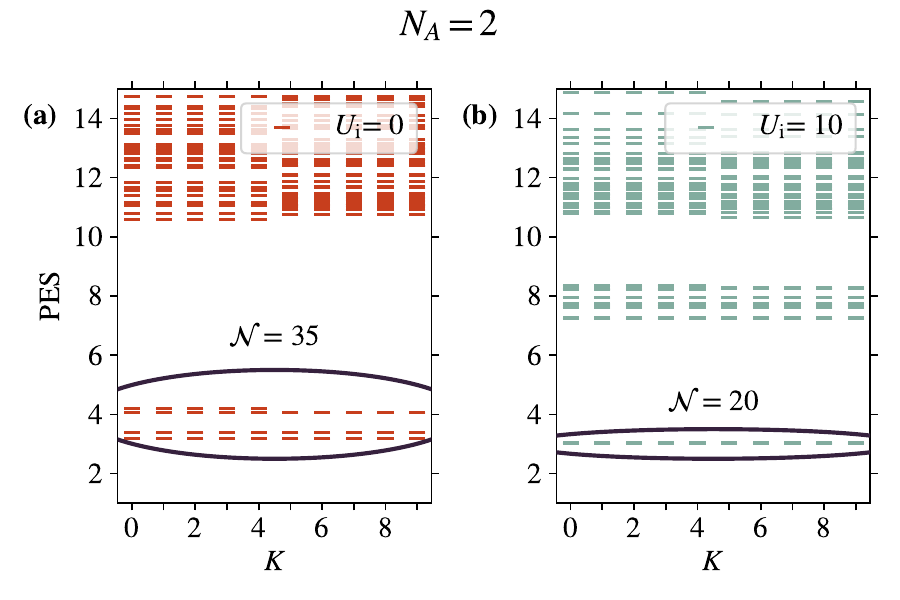}\hfill
    \includegraphics[width=0.8\linewidth]{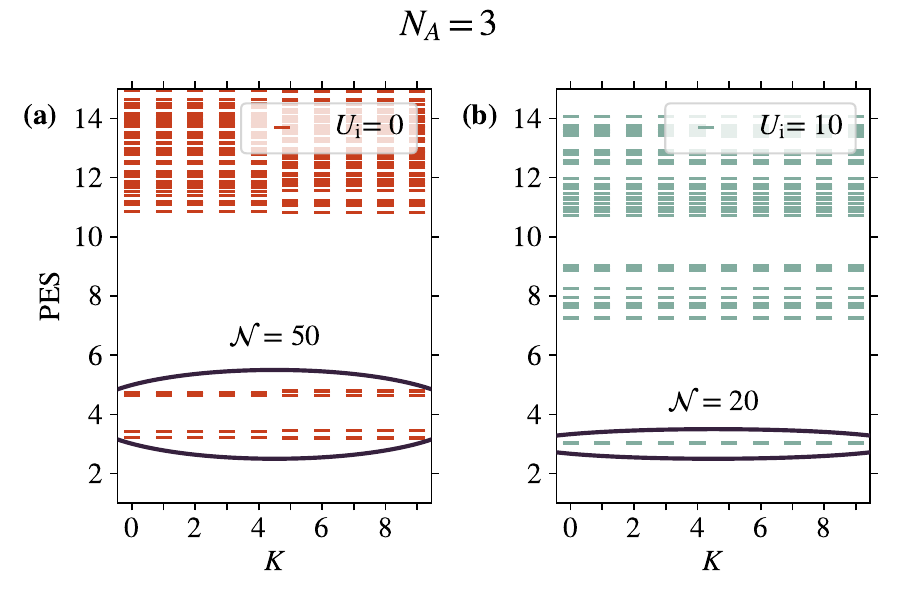}
    \caption{For a $10 \times 5$ lattice with $N=5$, there are 4 different particle bipartitions available which each produce different state countings.  We show here the other bipartitions to further prove that both the Laughlin, $U_{\mathrm{i}}=0$, case and the $k-DW$, $U_{\mathrm{i}} \neq 0$ case follow the expected counting.}
    \label{Fig:SuppMat:PES-all}
\end{figure}

\paragraph{\textbf{Other biparitions of the PES ---}}  We verify the state counting of both the trivial and topological regimes of our extended Harper-Hofstadter model by looking at other biparitions in the particle basis.  In Figure \ref{Fig:SuppMat:PES-all}, Eqs. \ref{Eq:PEScountingCDW} and \ref{Eq:PEScountingFCI} are verified by taking other particle bipartitions with $N_A=1,2,3$, and $4$.

\end{document}